\documentclass[%
 reprint,
superscriptaddress,
showkeys,
onecolumn,
amsmath,amssymb,
aps,
prf,
]{revtex4-2}

\usepackage{graphicx}
\usepackage{dcolumn}
\usepackage{bm}
\usepackage{soul}
\usepackage{hyperref}

\renewcommand{\vec}[1]{\boldsymbol{\bm{#1}}} 

\newcommand{\Ca}{\text{Ca}}
\newcommand{\Rey}{\text{Re}}
\newcommand{\gammadot}{\dot{\gamma}}

\newcommand{\Lx}{\text{L}_{\mbox{\scriptsize x}}}
\newcommand{\Ly}{\text{L}_{\mbox{\scriptsize y}}}
\newcommand{\Lz}{\text{L}_{\mbox{\scriptsize z}}}
\newcommand{\Bq}{\text{Bq}}

\newcommand{\taulb}{\tau_{\mbox{\tiny LB}}}

\newcommand{\muint}{\mu_{\mbox{\scriptsize int}}}

\newcommand{\mus}{\mu_{\mbox{\scriptsize s}}}
\newcommand{\mud}{\mu_{\mbox{\scriptsize d}}}

\newcommand{\Dst}{\text{D}_{\infty}}
\newcommand{\thetast}{\theta_{\infty}}
\newcommand{\Dtaylor}{\text{D}_{\mbox{\scriptsize T}}}
\newcommand{\Dshapira}{\text{D}_{\mbox{\scriptsize SH}}}

\newcommand{\Dflum}{\text{D}_{\mbox{\scriptsize F}}}

\newcommand{\Dinf}{\text{D}_{\infty}}

\newcommand{\thetaflum}{\theta_{\mbox{\scriptsize F}}}

\newcommand{\fmmone}{f_1^{(\mbox{\tiny MM})}}
\newcommand{\fmmtwo}{f_2^{(\mbox{\tiny MM})}}

\newcommand{\fMMi}{f_i^{(\mbox{\tiny MM})}}
\newcommand{\fNone}{f_1^{(\mbox{\tiny N})}}
\newcommand{\fNtwo}{f_2^{(\mbox{\tiny N})}}
\newcommand{\fNi}{f_i^{(\mbox{\tiny N})}}

\newcommand{\effi}{\mbox{n}_i}

\newcommand\corr[1]{\textcolor{black}{#1}}

\usepackage{xcolor}
\begin{document}

\preprint{APS/123-QED}


\title[Sample title]{
Analytical prediction for the steady-state behavior of a confined drop with interface viscosity under shear flow}

\author{F. Guglietta}
\email{fabio.guglietta@roma2.infn.it}
\affiliation{Department of Physics and INFN, Tor Vergata University of Rome, Via della Ricerca Scientifica 1, 00133 Rome, Italy}
\author{F. Pelusi}
\affiliation{Istituto per le Applicazioni del Calcolo - CNR, Via Pietro Castellino 111, 80131 Naples, Italy}

\date{\today}

\begin{abstract}
\corr{The steady-state behavior of a single drop under shear flow has been extensively investigated in the limit of small deformation and negligible inertia effects. In this work, we combine the calculations proposed by Flumerfelt [R. W. Flumerfelt, J. Colloid Interface Sci. {\bf 76}, 330 (1980)] for unconfined drops with interface viscosity, with those by Shapira and Haber [M. Shapira and S. Haber, Int. J. Multiph. Flow {\bf 16}, 305 (1990)] for confined drops without interface viscosity. By merging these two approaches, we provide comprehensive analytical predictions for steady-state drop deformation and inclination angle across a wide range of physical conditions, from confined to unconfined droplets, including or excluding the effect of interface viscosity. The proposed analytical predictions are also robust concerning variations in the viscosity ratio, making our model general enough to include any of the above conditions.}
\end{abstract}

\maketitle

\section{Introduction}
Complex fluids, such as blood, emulsions, and immiscible polymer blends, are familiar to people ordinary-life since they are implemented in several engineering applications, ranging from pharmaceutical~\cite{kim2004artificial,gao2015engineering,xia2019red} to petroleum~\cite{quintero2002overview,zolfaghari2016demulsification} and food industry~\cite{ofori2012use,tan2021application,udayakumar2021biopolymers}. During their processing, the micro-constituents of these systems, namely red blood cells, drops, and single polymers, may undergo deformation, turning into a specific system morphology, up to the interface rupture. Studying the steady-state shape of the system is crucial in defining the mechanical properties and rheology of such systems~\cite{guido1998three,narsimhan2019shape,aouaneStructureRheologySuspensions2021,Guglietta2023Suspension}. For this reason, precise control of the behavior of every single constituent under specific conditions is a critical aspect for enhancing and regulating manufacturing procedures. The most streamlined situation consists in a single drop undergoing a shear flow at low Reynolds numbers, which has garnered extensive scrutiny through analytical~\cite{taylor1932viscosity,taylor1934formation,chaffey1967second,rallison1980note,BarthesBieselAcrivos1973secondorder}, experimental~\cite{Torza1972experim,phillipsExperimentalStudiesDrop1980,grace1982dispersion}, and numerical investigations~\cite{guptaDeformationBreakupViscoelastic2014,komrakova2014lattice,megias2005droplet}.
In particular, under the same conditions, the pioneering work of Taylor~\cite{taylor1934formation} laid the foundation for predicting the steady-state behavior of both drop deformation and inclination angle with respect to the flow direction. However, when the drop is placed in confined geometries, the latter quantities suffer from significant variations regarding bulk systems~\cite{shapiraLowReynoldsNumber1990,sibilloDropDeformationMicroconfined2006,vananroyeEffectConfinementSteadystate2007,guido2011shear,vananroye2008microconfined}. 
In this scenario, Shapira \& Haber~\cite{shapiraLowReynoldsNumber1990} provided an analytical prediction for steady-state drop deformation as a function of the confinement degree and the relative distance between the drop's center of mass and the lower wall, but no claim has been made on the inclination angle. Their analytical prediction has also been confirmed by experiments~\cite{sibilloDropDeformationMicroconfined2006,vananroyeEffectConfinementSteadystate2007,guido2011shear}.
Later on, confinement degree has been numerically and experimentally observed to play an important role in drop breakup, influencing not only the threshold and dynamics of breakup but also the resulting daughter-drop size distribution under different confinement, viscosity ratios and flow conditions~\cite{janssen2010generalized,barai2016breakup,ioannou2016droplet,hernandez2017breakup, mulligan2011effect,anna2016droplets}.
Besides confinement effects, it has been detected that the presence of an interface viscosity represents an additional discriminating factor in determining the steady-state drop shape~\cite{flumerfelt1980effects,pozrikidisEffectsSurfaceViscosity1994,art:gounley16,art:lizhang19,narsimhan2019shape,guglietta2020effects}. Indeed, Flumerfelt~\cite{flumerfelt1980effects} extended Taylor's theory to account for the drop interface viscosity by providing analytical predictions for both drop deformation and inclination angle. In this landscape, a comprehensive analytical expression predicting the steady-state value of deformation and inclination angle of a confined drop with interface viscosity in shear flow is still missing, despite the necessity coming from practical applications involving drops and suspensions of droplets.

\begin{figure}[t!]
\centering
\includegraphics[width=1.\linewidth]{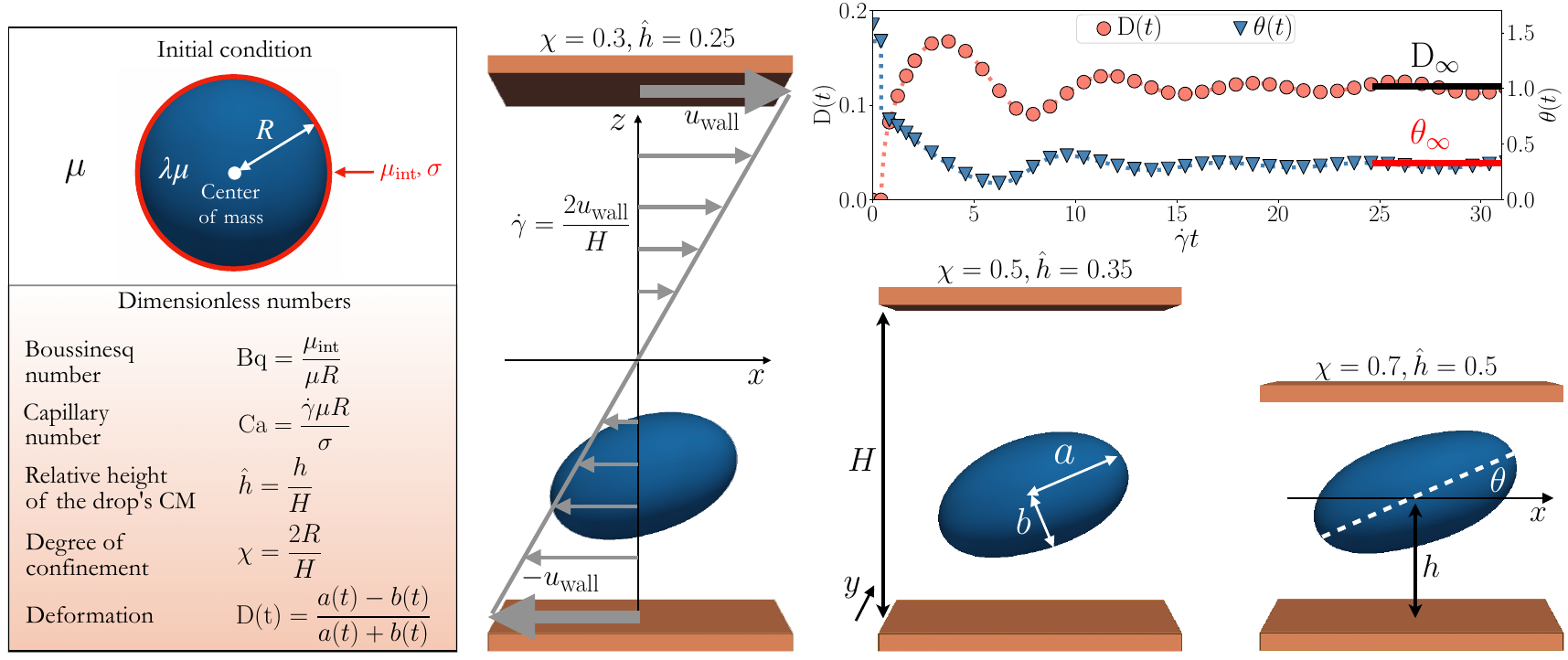}%
\caption{We consider a single drop with initial radius $R$, viscosity $\lambda \mu$ and surface tension $\sigma$ under shear flow, immersed in a fluid with viscosity $\mu$ and confined in a channel with two walls placed at distance $H$ moving with velocity $u_w(x,y,z = \pm H/2) = (\pm u_{\mathrm{wall}},0,0)$. The drop's center of mass is placed at a distance $h$ from the lower wall. We vary the Boussinesq ($\Bq$) and capillary ($\Ca$) numbers, the drop relative height ($\hat{h}$) and the degree of confinement ($\chi$). For each parameters' combination, we measure the time-evolution of both drop deformation $\mathrm{D}(t)$ and inclination angle $\theta(t)$ (see upper-right panel). The instantaneous deformation is computed as D$(t)=(a(t)-b(t))/(a(t)+b(t))$, where $a$ and $b$ are the drop main axes in the $xz$ plane. The steady-state value D$_\infty$ (solid black line) and $\theta_\infty$ (solid red line) are then considered. Data in the upper-right panel correspond to the situation with Ca$=0.33$, Bq$=40$, $\chi=0.7$, and $\hat{h}=0.5$.}\label{fig:sketch}%
\end{figure}

In this work, we present a \corr{comprehensive} analytical prediction for the steady-state deformation and inclination angle of a confined/unconfined drop with/without interface viscosity \corr{in the limit of small deformation and negligible inertia effects}.

\corr{The paper is organized as follows: in Sec.~\ref{sec:method}, both the employed numerical method and setup are presented. In Sec.~\ref{sec:deformation}, we report the derivation of the equation describing the steady-state deformation, while the derivation for the steady-state inclination angle is provided in Sec.~\ref{sec:angle}. Finally, conclusions are summarized in Sec.~\ref{sec:conclusions}.}

\section{Numerical method}\label{sec:method}

Given the inherent challenges of precisely controlling all possible parameter combinations in experiments is challenging, numerical simulations are crucial to address this scope.
In this work, we performed immersed boundary-lattice Boltzmann numerical simulations. Such a method has been largely benchmarked and employed for investigating the dynamics of single drops~\cite{art:lizhang19,guglietta2020effects,pelusi2023sharp,taglienti2023reduced,taglienti2024droplet} and capsules~\cite{zhangImmersedBoundaryLattice2007,kruger2011efficient,kruger2014interplay,Guglietta2023Suspension,silva2024lattice,owen2023lattice} with and without interface viscosity. In particular, the in-house GPU code we employed has already been extensively validated in previous works of some of the authors~\cite{guglietta2020effects,guglietta2021loading,guglietta2020lattice, taglienti2023reduced, taglienti2024droplet}.\newline

Instead of directly simulating the Navier-Stokes equations, the lattice Boltzmann (LB) method simulates the evolution of some probability distribution functions $\effi(\vec{x},t)$, representing the probability of finding fluid particles at position $\vec{x}$ and time $t$, moving with a discrete velocity $\vec{c}_i$. In particular, we implemented the so-called D3Q19 LB scheme, in which 19 discrete velocities are considered on a 3D regular lattice (see Ref.~\cite{book:kruger} for details). 
The dynamics of such functions $\effi$ is given by the LB equation:
\begin{equation}
    \effi(\vec{x}+\vec{c}_i\Delta t, t+\Delta t) - \effi(\vec{x}, t) = -\Delta t\frac{\effi(\vec{x}, t)-\effi^{\mbox{\tiny eq}}(\vec{x}, t)}{\taulb} + 
    \left(1-\frac{\Delta t}{2\taulb}\right)\frac{w_i}{c_s^2}\left[\left(\frac{\vec{c}_i\cdot\vec{u}}{c_s^2}+1\right)\vec{c}_i-\vec{u}\right]\cdot \vec{F}\ ,
\end{equation}
being $\effi^{\mbox{\tiny eq}}$ the equilibrium distribution function
\begin{equation}
\effi^{\mbox{\tiny eq}}(\vec{x},t)= w_i\rho\left(1+\frac{\vec{u}\cdot\vec{c}_i}{c_s^2}+\frac{(\vec{u}\cdot\vec{c}_i)^2}{2c_s^4}-\frac{\vec{u}\cdot\vec{u}}{\corr{2}c_s^2}\right)\ ,
\end{equation}
$\vec{F}$ the force density, $w_i$ suitable weights ($w_1=1/3$, $w_{2-7}=1/18$, $w_{8-19}=1/36$),  $c^2_s=1/3$ the speed of sound, $\Delta x=\Delta t= 1$ the lattice spacing and the discrete time-step, respectively, and $\taulb=1$ is the LB relaxation time~\cite{book:kruger}. 
The hydrodynamics fields (fluid density $\rho$ and velocity $\vec{u}$) can be computed as: 
\begin{equation}
     \rho(\vec{x},t) = \sum_{i=1}^{19}\effi(\vec{x},t)\ ,\qquad\qquad\qquad \rho\vec{u}(\vec{x},t) = \sum_{i=1}^{19}\vec{c}_i\effi(\vec{x},t)+\frac{\Delta t}{2}\vec{F}\ .
\end{equation}
Notice that $\taulb$ provides the link between the kinetic approach of the LB method and the hydrodynamic world of the Navier-Stokes equation via the fluid viscosity 
$\mu = \rho c_s^2\left(\taulb-\frac{\Delta t}{2}\right)$.
Further technical details can be found in textbooks~\cite{book:kruger,succi2018lattice} or in previous works by some of the authors, in which the same method (and the same numerical code) was employed~\cite{guglietta2020effects,guglietta2021loading,guglietta2020lattice,taglienti2023reduced,pelusi2023sharp,taglienti2024droplet}.
In the present work, the fluid domain is a three-dimensional rectangular box with sizes $\Lx=\Ly=128 \ \Delta x$ and $\Lz=H$, the latter varying to explore different degrees of confinement.
The immersed boundary method handles the two-way coupling between the fluid and the drop interface~\cite{book:kruger,peskin2002immersed}. 
The drop interface is modeled as a 3D triangular mesh with radius $R=20 \ \Delta x$, made of $1.8\times 10^4$ faces. On each mesh Lagrangian node $k$ at position $\vec{r}_k$, the interface force $\vec{\varphi}_k(t)$ is computed (details on the interface model are given below), and then, according to the immersed boundary method, it is spread over the neighboring Eulerian (fluid, LB) nodes with coordinates $\vec{x}$: 
\begin{equation}
\vec{F}(\vec{x},t) = \sum_k \vec{\varphi}_k(t)\delta(\vec{r}_k-\vec{x})\ .
\end{equation} 
The function $\delta(\vec{x})$ is the discretized Dirac delta. In particular, we implemented the 4-points stencil for the Dirac delta (see Refs.~\cite{book:kruger,peskin2002immersed}). The immersed boundary method is also used to interpolate the fluid velocity on the $k-$th Lagrangian node: 
\begin{equation}
\vec{\dot{r}}_k(t) = \sum_{\vec{x}}\Delta x^3\vec{u}(\vec{x},t)\delta(\vec{r}_k-\vec{x})\ .    
\end{equation} 
Once we have the velocity $\vec{\dot{r}}_k$, the position of the Lagrangian node is integrated with a forward-Euler scheme.\newline

Concerning the interface model, we compute on each element of the triangular mesh the viscoelastic stress tensor $\vec{\tau} = \vec{\tau}_s+\vec{\tau}_v$, that is made of the sum of the surface-tension stress tensor $\vec{\tau}_s=\sigma \mathcal{I}_2$~\cite{art:lizhang19,guglietta2020effects,taglienti2023reduced,taglienti2024droplet}, with $\mathcal{I}_2$ being the 2D identity matrix, and the viscous tensor $\vec{\tau}_v$, which is given by the Boussinesq-Scriven tensor~\cite{Scriven1960}: \corr{$\vec{\tau}_v = 2\muint\vec{e}$, where  $\vec{e}=\frac{1}{2}[\nabla\vec{u}+(\nabla\vec{u})^T]$ is the strain rate tensor and} $\nabla\vec{u}$ is the velocity gradient at the interface~\cite{art:lizhang19,barthesbiesel1985,art:gounley16}. 
\corr{In the most general case, the Boussinesq-Scriven tensor can be split into shear and dilatational contributions, characterized by the corresponding values of surface viscosity ($\mus$ and $\mud$, respectively): $\vec{\tau}_v = \mus[2\vec{e}-\mbox{tr}(\vec{e})\mathcal{I}_2]+\mud\mbox{tr}(\vec{e})\mathcal{I}_2$. 
For the sake of simplicity, as already done in other works~\cite{barthesbiesel1985,art:yazdanibagchi13}, we assume the surface shear and dilatational viscosities to be the same, allowing us to characterize viscous effects with a single parameter: $\mus=\mud=\muint$.}
Numerical details on how to compute the nodal force $\vec{\varphi}_i$ starting from the stress tensor $\vec{\tau}$ can be found in Refs.~\cite{art:lizhang19,guglietta2020effects}.\newline

\corr{Concerning the setup,} we consider a drop with initial radius $R$ characterized by a surface tension $\sigma$, an interface viscosity $\muint$, and immersed in a fluid with density $\rho$ and viscosity $\mu$, resulting in a viscosity ratio between inner and outer fluid viscosity $\lambda$ (see Fig.~\ref{fig:sketch}). The drop is placed between two walls in the $xy$ plane, located at $z=\pm H/2$ and separated by a distance $H$. The drop's center of mass is fixed at a distance $h$ from the lower wall.
We apply a shear rate $\gammadot = 2 u_{\mathrm{wall}}/H$ by moving the walls with a velocity $u_w(x,y,z=\pm H/2) = (\pm u_{\mathrm{wall}},0,0)$ (see Fig.~\ref{fig:sketch}). The Reynolds number $\Rey=\rho\gammadot R^2/\mu$ keeps values smaller than $10^{-2}$ to avoid inertial effects. We measure the drop steady-state deformation $\Dinf$ and the inclination angle $\theta_\infty$ by varying the capillary number $\Ca=\gammadot\mu R/\sigma$, the Boussinesq number $\Bq=\muint/\mu R$, the degree of confinement $\chi = 2R/H$, and the relative height of the drop's center of mass $\hat{h}=h/H$.\newline

\section{Steady-state deformation}\label{sec:deformation}
\begin{figure}[t!]
\centering
\includegraphics[width=.85\linewidth]{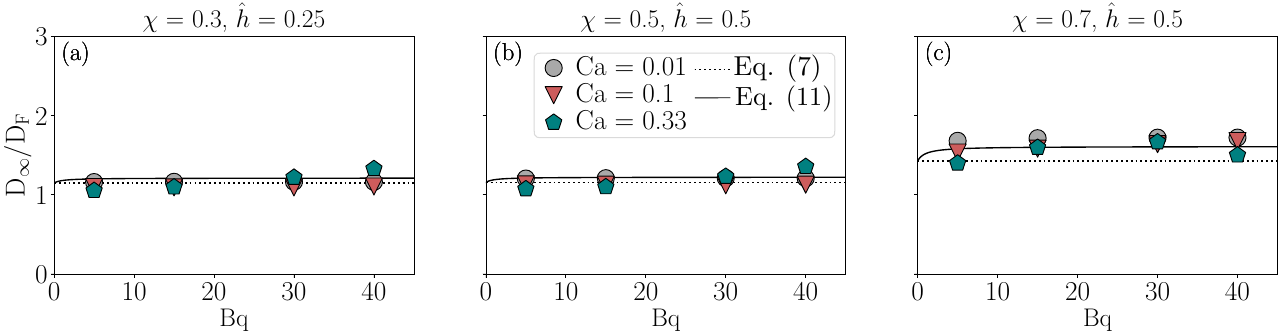}%
\caption{For three representative combinations of $\chi$ and $\hat{h}$ with $\lambda =1$, we show data for the measured steady-state value of the drop deformation D$_\infty$, normalized with the corresponding value of D$_{\mathrm{F}}$ (cfr. Eq.~\eqref{eq:def_flum}), as a function of Bq. Dotted \corr{and solid black lines report the corresponding value of the function $\Psi_{\corr{\mathrm{SH}}}$ (cfr. Eq.~\eqref{eq:psi1}) and $\Psi$ (dfr. Eq.~\eqref{eq:psi_oldroyd}), respectively.} Different colors/symbols correspond to different values of Ca.}\label{fig:D_ratio_with_flumerfelt}%
\end{figure}

To derive an analytical prediction for the steady-state drop deformation under shear flow, we considered the one computed by Shapira \& Haber~\cite{shapiraLowReynoldsNumber1990} for confined pure drops, which reads 
\begin{equation}\label{eq:SH}
\Dshapira(\lambda,\hat{h},\chi,\Ca)=\Dtaylor(\lambda,\Ca)\Psi_{\corr{\mathrm{SH}}}(\lambda,\hat{h},\chi) \ ,
\end{equation}
where $\Dtaylor(\lambda,\Ca)=\Ca(19\lambda+16)/(16\lambda+16)$ is the steady-state drop deformation computed by Taylor for an unconfined pure drop~\cite{taylor1932viscosity}, and 
\begin{equation}\label{eq:psi1}
\Psi_{\corr{\mathrm{SH}}}(\lambda,\hat{h},\chi)=1+C_s(\hat{h})\left(\frac{\chi}{2}\right)^3\frac{1+2.5\lambda}{1+\lambda}\ ,
\end{equation}
is a function accounting for the degree of confinement. The coefficient $C_s(\hat{h})$ in Eq.~\eqref{eq:psi1} represents the shape factor~\footnote{The values of $C_s(\hat{h})$ used in this work are $C_s(\hat{h}=0.25) = 25.026798$, $C_s(\hat{h}=0.35) = 9.9646144$ and $C_s(\hat{h}=0.5) = 5.6996174$ and are taken from Ref.~\cite{shapiraLowReynoldsNumber1990}}. It is worth noting that, when $\chi=0$ (unconfined drop), then $\Psi_{\corr{\mathrm{SH}}}=1$ and $\Dshapira=\Dtaylor$. To include the effect of the interface viscosity, we considered the steady-state deformation of an unconfined drop with interface viscosity computed by Flumerfelt~\cite{flumerfelt1980effects}:
\begin{equation}\label{eq:def_flum}
 \Dflum(\lambda,\Ca,\Bq) = \frac{19\lambda+16+32\Bq}{(16\lambda+16+32\Bq)\sqrt{\Ca^{-2} +\left[\frac{19 \mathcal{F}}{20}(\lambda+2\Bq)\right]^{\corr{2}}}} \ ,
\end{equation}
with 
\begin{equation}\label{eq:F_flum}
    \mathcal{F}=1- \frac{9\lambda+18\Bq-2}{8(\lambda+2\Bq)^2} \ .
\end{equation}
Based on this observation and intending to provide a \corr{comprehensive} analytical prediction for confined droplets with interface viscosity, we replaced $\Dtaylor(\lambda,\Ca)$ with $\Dflum (\lambda,\Ca,\Bq)$ in Eq.~\eqref{eq:SH}, thus obtaining 
\begin{equation}\label{eq:D_short}
\mbox{D}(\lambda,\hat{h},\chi,\Ca,\Bq) = \Dflum (\lambda,\Ca,\Bq) \Psi_{\corr{\mathrm{SH}}} (\lambda,\hat{h},\chi) \ .
\end{equation}
In the case of $\chi=0$ (i.e., $\Psi_{\corr{\mathrm{SH}}}=1$), Eq.~\eqref{eq:D_short} recovers both Flumerfelt's result (when $\Bq>0$) and Taylor's result (when $\Bq=0$).
In this way, we implicitly assume that $\Psi_{\corr{\mathrm{SH}}}(\lambda,\hat{h},\chi)$ does not depend on the interface viscosity and that the effects of confinement and interface viscosity separately contribute to the overall drop deformation. \corr{On the other hand, one can consider the Oldroyd's model~\cite{oldroyd1955effect} to account for the effect of membrane viscosity, as also shown in Refs.~\cite{danov2001viscosity,narsimhan2019shape}. 
This results in considering an effective viscosity ratio $\lambda^*=\lambda+2\Bq$~\cite{danov2001viscosity,narsimhan2019shape,flumerfelt1980effects,art:gounley16}, that can be substituted in Eq.~\eqref{eq:psi1}, leading to:
\begin{equation}\label{eq:psi_oldroyd}
\Psi(\lambda,\hat{h},\chi, \Bq)=1+C_s(\hat{h})\left(\frac{\chi}{2}\right)^3\frac{1+2.5\lambda + 5 \Bq}{1+\lambda + 2 \Bq} \ .
\end{equation}
However, Eq.~\eqref{eq:psi_oldroyd} represents a small correction to Eq.~\eqref{eq:psi1}, which can be only appreciated for high degrees of confinement. To verify it}, we measured the steady-state value of the deformation $\Dinf$ from numerical simulations, and we estimated the ratio $\Dinf/\Dflum$ as a function of $\Bq$. Results are reported in Fig.~\ref{fig:D_ratio_with_flumerfelt} for $\lambda=1$: different panels refer to different combinations of $\chi$ and $\hat{h}$, while different symbols and colors refer to different $\Ca$. In each panel, \corr{the corresponding values of $\Psi_{\corr{\mathrm{SH}}}(\lambda,\hat{h},\chi)$ (cfr. Eq.~\eqref{eq:psi1}) and $\Psi(\lambda,\hat{h},\chi, \Bq)$ (cfr. Eq.~\eqref{eq:psi_oldroyd}) are reported with dashed and solid black lines, respectively. Our analysis shows that at small degrees of confinement ($\chi < 0.7$), $\Phi_{\mathrm{SH}}$ and $\Psi$ are qualitatively the same, with $\Psi$ losing its dependence on $\Bq$ for values of $\Bq > 5$; however, when $\chi$ is large, then a discrepancy emerges.} 

By substituting the expression for $\Dflum$ (cfr. Eq.~\eqref{eq:def_flum}) and that for \corr{$\Psi$ (cfr. Eq.~\eqref{eq:psi_oldroyd})} in Eq.~\eqref{eq:D_short}, we obtained the explicit \corr{comprehensive} analytical expression for the steady-state deformation of a confined drop with interface viscosity:
\begin{equation}\label{eq:def}
\mbox{D}(\lambda,\hat{h},\chi,\Ca,\Bq) = \frac{19\lambda+16+32\Bq}{(16\lambda+16+32\Bq)\sqrt{\Ca^{-2} +\left[\frac{19 \mathcal{F}}{20}(\lambda+2\Bq) \right]^{\corr{2}}}}\left[1+C_s(\hat{h})\frac{\chi^3}{8}\frac{1+2.5\lambda \corr{+ 5 \Bq}}{1+\lambda\corr{+ 2 \Bq}}\right] \ . 
\end{equation}
We remark that in the two limit cases of an unconfined drop with interface viscosity ($\chi=0$, $\Bq>0$) and a confined pure drop ($\chi>0$, $\Bq=0$), Eq.~\eqref{eq:def} exactly recovers Flumerfelt's and Shapira \& Haber's predictions, respectively.

To further validate Eq.~\eqref{eq:def}, we compared analytical prediction with numerical data. Since drop deformation in Eq.\eqref{eq:def} depends on five parameters ($\lambda$, $\hat{h}$, $\chi$, $\Ca$, $\Bq$), we started by fixing $\lambda=1$ and varying the remaining parameters. We do not show data in either the case of an unconfined drop with interface viscosity or the confined pure drop because results have already been widely explored~\cite{flumerfelt1980effects,shapiraLowReynoldsNumber1990}. 

\begin{figure}[t!]
\centering
\includegraphics[width=.95\linewidth]{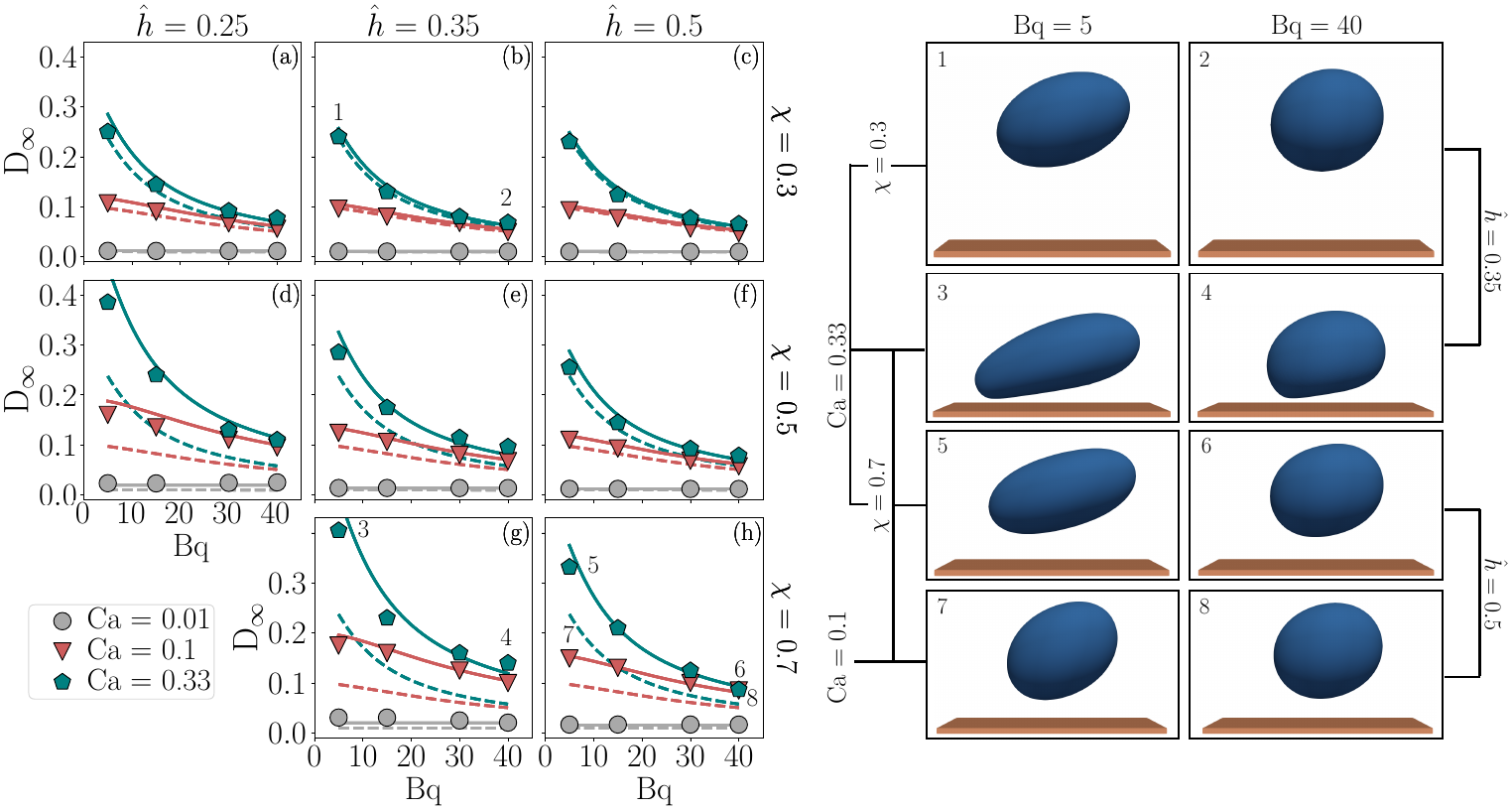}
\caption{Left panels: Measured steady-state value of the drop deformation D$_\infty$ as a function of Bq for $\lambda =1$. Symbols show simulation data, while solid and dotted lines correspond to Eq.~\eqref{eq:def} and Flumerfelt's prediction $\Dflum$ (cfr. Eq.~\eqref{eq:def_flum}), respectively. Different colors/symbols correspond to different values of Ca. Right panels: steady-state shape of droplets for different parameters' combination that are marked in left panels with index $i=1,\dots,8$.\label{fig:deformation_vs_bq}}%
\end{figure}

Fig.~\ref{fig:deformation_vs_bq} reports the measured steady-state deformation $\Dst$ as a function of $\Bq$ for different drop relative height $\hat{h}$ (columns) and various degrees of confinement $\chi$ (rows). Symbols represent numerical simulation data, while solid lines show the theoretical prediction given by Eq.~\eqref{eq:def}. Different colors refer to different values of $\Ca$.
The overall effect of increasing confinement is to enhance the drop deformation, as already known for the pure drop~\cite{shapiraLowReynoldsNumber1990}; on the other hand, the effect of interface viscosity is to reduce the deformation~\cite{flumerfelt1980effects}. The combination of both effects holds the same qualitative scenario. Numerical results and analytical predictions are in good agreement \corr{, especially in the limit of small $\Ca$ and $\chi$, which is the limit in which the calculations by Flumerfelt and Shapira \& Haber were performed. There is, indeed,} a slight difference for the combination of strong confinement ($\chi= 0.7$), large values of interface viscosity ($\Bq\ge 30$), and low relative drop height ($\hat{h}=0.35$). Such differences have also been observed experimentally for pure drops~\cite{sibilloDropDeformationMicroconfined2006,minale2008phenomenological}, where experiments for high values of capillary number ($\Ca\approx 0.3$) and high degrees of confinement ($\chi>0.5$) show that the prediction by Shapira \& Haber underestimates the measured values. Notice that simulation data for the case $\chi=0.7$ and $\hat{h}=0.25$ does not exist because the distance $h$ would be smaller than $R$ (see Fig.~\ref{fig:sketch}). Further, we remark that the correction that our model provides on the already existing theories is pronounced: indeed, dashed lines in Fig.~\ref{fig:deformation_vs_bq} refer to Flumerfelt's prediction $\Dflum$ (cfr. Eq.~\eqref{eq:def_flum}) for the unconfined drop with interface viscosity. As expected, Eq.~\eqref{eq:def} approaches $\Dflum$ only for small values of $\chi$. 
To visually capture the effect of both interface viscosity and confinement on drop shape, right panels (1-8) of Fig.~\ref{fig:deformation_vs_bq} show snapshots of the steady-state configurations for selected and relevant combinations of $\Ca, \hat{h}$ and $\chi$, spanning from the less confined drop with a small interface viscosity (panel (1)) to the most confined case with the highest considered value of $\Bq$ (panel (8)).
The effect of interface viscosity can be appreciated by comparing left ($\Bq=5$) and right ($\Bq=40$) panels. In contrast, the impact of the proximity to the wall is reflected in a not-symmetric drop shape (panels (3-4)). This asymmetry is mitigated by the impact of interface viscosity.\newline

\section{Steady-state inclination angle}\label{sec:angle}
We now move our attention to the drop inclination angle with respect to the flow direction. According to Maffettone \& Minale (MM)~\cite{maffettone1998equation}, who described the shape evolution of an ellipsoidal drop in terms of 
two functions $f_1$ and $f_2$, the steady-state deformation and inclination angle in the limit of small $\Ca$ can be computed as 
\begin{subequations}
\begin{align}
&\mbox{D} =\Ca \frac{f_2}{2f_1}\ , \label{eq:D_lin}\\
&\theta =\frac{1}{2}\arctan{\left(\frac{f_1}{\Ca}\right)}\ , \label{eq:theta_lin}
\end{align}
\end{subequations}
respectively.
Since they considered an unconfined pure drop ($\chi=0$ and $\Bq=0$),
the functions $f_i$ have been chosen to recover the first-order expansion in $\Ca$ of the shape-evolution equation computed by Rallison~\cite{rallison1980note}: 
\begin{subequations}
\begin{align}
&f_1(\lambda)=\fmmone(\lambda)=\frac{40(\lambda+1)}{(2\lambda+3)(19\lambda+16)}\ ,\\
&f_2(\lambda)=\fmmtwo(\lambda)=\frac{5}{2\lambda+3}\ .
\end{align}
\end{subequations}
Since Rallison's result has been recently extended by Narsimhan (N)~\cite{narsimhan2019shape} to include the effect of the interface viscosity, we matched the functions $f_i$ with the first-order expansion of the shape-evolution equation computed in Ref.~\cite{narsimhan2019shape}, thus obtaining $f_i(\lambda,\Bq)=\fNi(\lambda,\Bq)$ (cfr. Eqs.~\eqref{eq:fnars} in Appendix~\ref{app:f1f2}). Notice that, in the limit of pure drop, $\fNi(\lambda,\Bq=0)$ coincide with $\fMMi(\lambda)$. To include the effect of the confinement degree, we followed Minale's approach~\cite{minale2008phenomenological}, who extended $\fMMi$ to account for the effect of $\chi$. Thus, we extended $f_i$ as
\begin{subequations}
\begin{align}
&f_1^{\mathrm{ext}} = \fNone(\lambda,\Bq) / \psi_1 (\lambda,\hat{h},\chi)\ ,\\
&f_2^{\mathrm{ext}} = \fNtwo(\lambda,\Bq) \psi_2 (\lambda,\hat{h},\chi)\ ,
\end{align}
\end{subequations}
with 
\begin{equation}
\psi_i=1+C_s(\hat{h})\frac{\chi^3}{8}\tilde{f}_i\ ,
\end{equation}
where the expressions for $\tilde{f}_i$ have to be matched with the analytical results by Shapira \& Haber~\cite{shapiraLowReynoldsNumber1990} (in Ref.~\cite{minale2008phenomenological}, $\fNi$ are replaced by $\fMMi$ in the definition of $f_i^{\mathrm{ext}}$). 
To obtain an expression for $\tilde{f}_i$, we substituted $f_i$ with $f_i^{\mathrm{ext}}$ in Eq.~\eqref{eq:D_lin} and, by neglecting $\mathcal{O}(\chi^6)$, we obtained:
\begin{equation}\label{eq:psi}
\mbox{D}(\lambda,\hat{h},\chi,\Ca,\Bq)=\Ca \frac{\fNtwo}{2\fNone} \psi_1\psi_2 =\Ca \frac{19\lambda+16+32\Bq}{16\lambda+16+32\Bq}\left[1+C_s(\hat{h})\frac{\chi^3}{8}(\tilde{f}_1+\tilde{f}_2)\right] \ .
\end{equation}

We then considered the first-order expansion in $\Ca$ of Eq.~\eqref{eq:def} \corr{with the assumption $\Psi = \Psi_{\mathrm{SH}}$}, i.e., 
\begin{equation}\label{eq:def_first}
    \mbox{D}(\lambda,\hat{h},\chi,\Ca,\Bq)=\Ca \frac{19\lambda+16+32\Bq}{16\lambda+16+32\Bq}\left[1+C_s(\hat{h})\frac{\chi^3}{8}\frac{1+2.5\lambda}{1+\lambda}\right]\ .
\end{equation}
\corr{The assumption that $\Psi = \Psi_{\mathrm{SH}}$ follows from Shapira \& Haber's calculations~\cite{shapiraLowReynoldsNumber1990}: indeed, the last hold for small values of $\chi$ and, as we showed in Fig.~\ref{fig:D_ratio_with_flumerfelt}, $\Psi \approx \Psi_{\mathrm{SH}}$ for small degrees of confinement (i.e., $\chi<0.7$).}

By comparing Eq.~\eqref{eq:psi} with Eq.~\eqref{eq:def_first} it is straightforward to see that
\begin{equation}
\tilde{f}_1+\tilde{f}_2=\frac{1+2.5\lambda}{1+\lambda}\ . 
\end{equation}
This implies that, once $\tilde{f}_2$ is determined, $\tilde{f}_1$ is fixed. 
\corr{Since} $\Psi_{\corr{\mathrm{SH}}}(\lambda,\hat{h},\chi)$ (i.e., the term in squared parenthesis in Eq.~\eqref{eq:def_first}) does not depend on $\Bq$, \corr{then} neither $\tilde{f}_1$ nor $\tilde{f}_2$ depend on $\Bq$. \corr{Therefore,} the values computed by Minale~\cite{minale2008phenomenological} for a pure confined drop are still valid:
\begin{subequations}\label{eq:f1f2}
\begin{align}
    \tilde{f_1}(\lambda) &= 
    \frac{22+32\lambda-\frac{13}{2}\lambda^2}{12+13\lambda+\lambda^2}\ , \\ 
    \tilde{f}_2(\lambda) &= - \frac{10-9\lambda}{12+\lambda}\ .
\end{align}
\end{subequations}

\begin{figure}[t!]
\centering
\includegraphics[width=.51\linewidth]{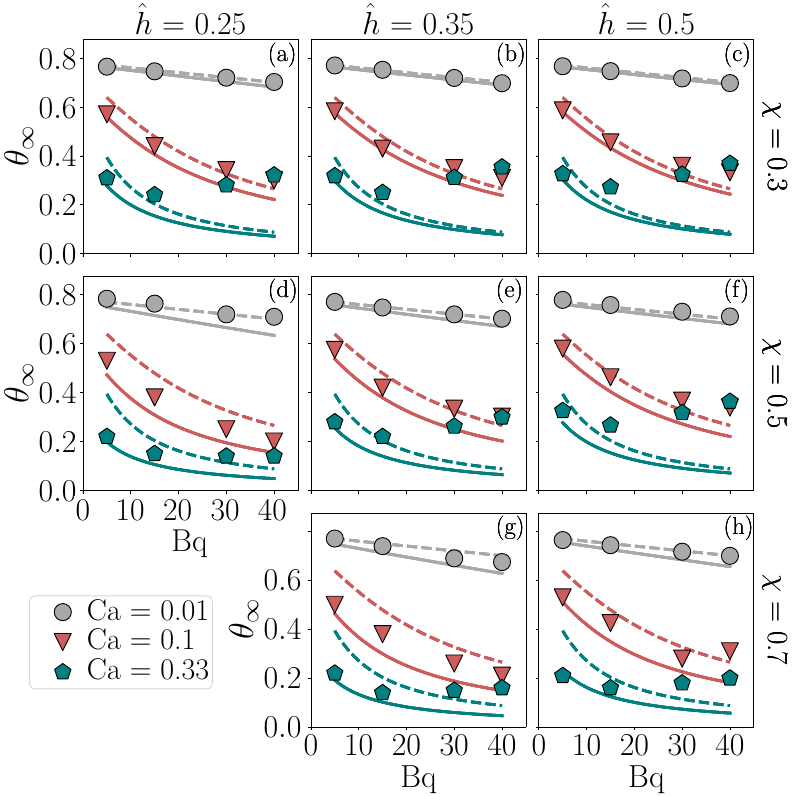}%
\caption{Steady-state value of the drop inclination angle $\theta_\infty$ as a function of Bq for $\lambda =1$. 
Symbols show simulation data, while solid and dotted lines draw Eq.~\eqref{eq:angle} and Flumerfelt's prediction $\thetaflum$ (cfr. Eq.~\eqref{eq:theta_flum}), respectively. Different colors/symbols correspond to different values of Ca.}\label{fig:angle_vs_bq}%
\end{figure}
Therefore, the prediction for the steady-state inclination angle is given by Eq.~\eqref{eq:theta_lin} with $f_1 = f_1^{\mathrm{ext}}=\fNone/\psi_1$:
\begin{equation}\label{eq:angle}
\theta(\lambda,\hat{h},\chi,\Ca,\Bq) =\frac{1}{2}\arctan\left(\frac{40\left(\lambda+1\right)+80\Bq}{\left(2\lambda+3\right)\left(19\lambda+16\right)+\Bq(32\Bq+98\lambda+112)}\frac{1}{\left[1+C_s(\hat{h})\frac{\chi^3}{8}\tilde{f}_1(\lambda)\right]\Ca}\right)\ .
\end{equation}
When $\Bq=0$, Minale's results are retrieved~\cite{minale2008phenomenological}.

In Fig.~\ref{fig:angle_vs_bq}, we compare Eq.~\eqref{eq:angle} (solid lines) with  Flumerfelt's result (cfr. Eq.~\eqref{eq:theta_flum}, dashed lines), showing that accounting for the effect of confinement in the prediction for the steady-state inclination angle leads to a significant difference.
Numerical simulations (symbols) and Eq.~\eqref{eq:angle} are in good agreement, showing a slight mismatch when $\Ca$ increases or when the drop is close to the wall. 
This mismatch is not surprising for two reasons: first,  Eq.~\eqref{eq:angle} has been derived in the limit of small $\Ca$; second, Eq.~\eqref{eq:angle} comes from Maffettone \& Minale model, in which the drop is assumed to be always ellipsoidal, whereas simulations show a more complex shape, especially in proximity to the wall (cfr. Fig.~\ref{fig:deformation_vs_bq}, right panels). 
Numerical simulations also reveal a non-monotonic behavior of $\theta$ as $\Bq$ increases, which has already been observed in other works concerning both unconfined drops~\cite{guglietta2020effects} and unconfined capsules~\cite{art:yazdanibagchi13} with interface viscosity. 
Finally, we observed that the proximity of the drop to the wall influences the inclination angle measurement. In such cases, the drop shape loses symmetry relative to the largest diameter in the shear plane (see Fig.~\ref{fig:deformation_vs_bq}, right panels). 
Notice that we measured the inclination angle considering the segment connecting the drop's center of mass to the top-right part of the shape (as shown by the white segment of length $a$ in Fig.~\ref{fig:sketch}) and the flow direction.

\begin{figure}[t!]
\centering
\includegraphics[width=.75\linewidth]{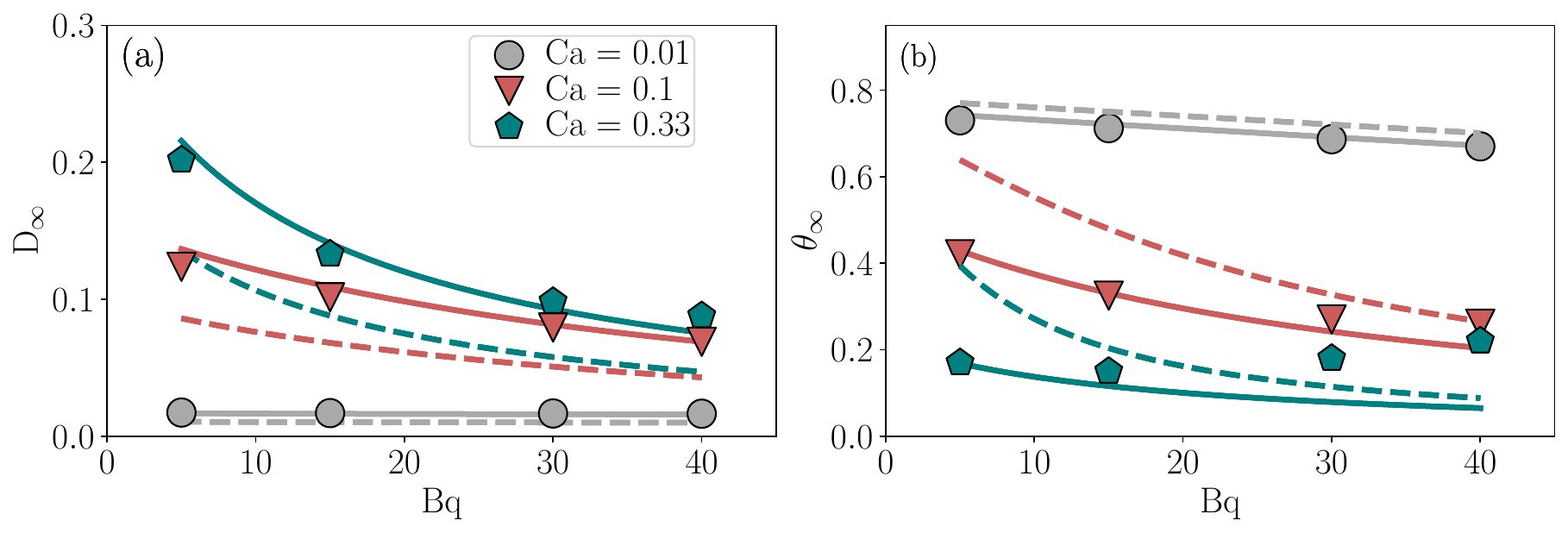}%
\caption{Steady-state value of the drop deformation D$_\infty$ (panel (a)) and inclination angle $\theta_\infty$ (panel (b)) as a function Bq for a value of $\lambda = 5.2$, $\chi=0.7$, and $\hat{h}=0.5$. Symbols show simulation data, while solid lines correspond to predictions from Eq.~\eqref{eq:def} (panel (a)) and Eq.~\eqref{eq:angle} (panel (b)). 
Dashed lines draw Flumerfelt's predictions, i.e. Eq.~\eqref{eq:def_flum} (panel (a)) and  Eq.~\eqref{eq:theta_flum} (panel (b)), respectively.
Different colors/symbols correspond to different values of Ca.}\label{fig:deformation_vs_bq_lambdas}%
\end{figure}

All results hitherto discussed refer to the case of $\lambda=1$. To prove the robustness of the analytical expressions for both D (cfr. Eq.~\eqref{eq:def}) and $\theta$ (cfr. Eq.~\eqref{eq:angle}), we also explored a situation with $\lambda = 5.2$, which is usually encountered in experiments~\cite{minale2008phenomenological,vananroyeEffectConfinementSteadystate2007}.
In Fig.~\ref{fig:deformation_vs_bq_lambdas}, $\Dst$ (panel (a)) and $\thetast$ (panel (b)) are reported as functions of $\Bq$ for the most confined simulated case (i.e., $\chi=0.7$) and with $\hat{h}=0.5$. In contrast with $\Dst$, which shows a perfect agreement between simulation data and Eq.~\eqref{eq:def}, $\thetast$ behaves as in Fig.~\ref{fig:angle_vs_bq}, with a good deal with Eq.~\eqref{eq:angle} for small values of $\Ca$ and a disagreement for $\Ca=0.33$. This result further confirms the validity of our analytical predictions. \newline

\section{Conclusions}\label{sec:conclusions}
In this work, \corr{we present a \corr{comprehensive} analytical prediction describing the steady-state deformation (cfr. Eq.~\eqref{eq:def}) and inclination angle (cfr. Eq.~\eqref{eq:angle}) of a confined/unconfined drop with/without interface viscosity, in the limit of small deformation and negligible inertia effects.}
Since there are no available experiments for this system, to validate the quality and robustness of the proposed predictions, we performed immersed boundary-lattice Boltzmann simulations, which have already been validated in a series of previous works~\cite{art:lizhang19,guglietta2020effects,taglienti2023reduced,taglienti2024droplet,pelusi2023sharp}. We also compared Eqs.~\eqref{eq:def} and \eqref{eq:angle} with the already known analytical predictions computed by Flumerfelt~\cite{flumerfelt1980effects} for the unconfined drop with interface viscosity, showing that the accounting for confinement effects in Flumerfelt's prediction is fundamental to quantitatively describe the correct deformation of a drop under these conditions. The robustness of our analytical predictions is strengthened by its application to different values of the drop viscosity ratio. We hope this result can open a new route to experiments of confined/unconfined drops with/without interface viscosity.\newline 

\section{Acknowledgment} 
We thank L. Biferale and M. Sbragaglia for fruitful discussions. This work was supported by the Italian Ministry of University and Research (MUR) under the FARE program (No. R2045J8XAW), project ``Smart-HEART''. FP acknowledges the support of the National Center for HPC, Big Data and Quantum Computing, Project CN\_00000013 – CUP B93C22000620006, Mission 4 Component 2 Investment 1.4, funded by the European Union – NextGenerationEU.

\appendix
\section*{APPENDIX}

\subsection{Functions $\fNone$ and $\fNtwo$}\label{app:f1f2}
To include the interface viscosity, similarly to what Maffettone \& Minale~\cite{maffettone1998equation} did with Rallison's result~\cite{rallison1980note} for the pure (i.e., $\Bq=0$) unconfined drop, one can match the values of $f_{1,2}$ with the first-order expansion of Eq.(3.1) in Ref.~\cite{narsimhan2019shape}, thus obtaining:
\begin{subequations}\label{eq:fnars}\begin{align}
    \fNone &= \frac{40\left(\lambda+1\right)+80\Bq}{\left(2\lambda+3\right)\left(19\lambda+16\right)+\Bq(32\Bq+98\lambda+112)}\ , \\
    \fNtwo &= 5\frac{19\lambda+16+32\Bq}{\left(2\lambda+3\right)\left(19\lambda+16\right)+\Bq(32\Bq+98\lambda+112)}\ .
\end{align}\end{subequations}

\subsection{Flumerfelt's prediction for inclination angle}\label{app:flum}
The steady-state inclination angle $\thetaflum$ with respect to the flow direction which has been predicted by Flumerfelt~\cite{flumerfelt1980effects} is:
\begin{equation}\label{eq:theta_flum}
\thetaflum(\lambda,\Ca,\Bq) =  \frac{1}{2}\arctan\left(
\frac{20}{19\mathcal{F} (\lambda+2\Bq)\Ca }
\right)\ ,
\end{equation}
where $\mathcal{F}$ is given in Eq.~\eqref{eq:F_flum}.

\bibliography{main}

\end{document}